\documentclass[a4paper,11pt]{article}
\pdfoutput=1 

\usepackage{jheppub} 
\usepackage{xspace}
\usepackage{hepunits}
\usepackage{booktabs}
\usepackage[utf8]{inputenc}                    
\usepackage[T1]{fontenc} 
\usepackage{amsmath}

\newcommand{\seff}{\ensuremath{\sigma_{\textit{eff}}}\xspace}
\newcommand{\seffcdf}{\ensuremath{\sigma_{\textit{eff,CDF}}}\xspace}

\newcommand{\seffd}{\ensuremath{\sigma_{\textit{eff,D0}}}\xspace}
\newcommand{\seffdcdf}{\ensuremath{\sigma_{\textit{eff,D0+reCDF}}}\xspace}
\newcommand{\seffcdfre}{\ensuremath{\sigma_{\textit{eff,reCDF}}}\xspace}
\newcommand{\seffmodel}{\ensuremath{\sigma_{\textit{eff,model}}}\xspace}

\newcommand{\herwig}{\mbox{\textsc{Herwig}}\xspace}
\newcommand{\pythia}{\textsc{Pythia}\xspace}

\newcommand{\Professor}{\textsc{Professor}\xspace}

\newcommand{\ptmin}{\ensuremath{p_{\perp}^{\rm min}}\xspace}
\newcommand{\pdisrupt}{\ensuremath{p_{\rm disrupt}}\xspace}
\newcommand{\dNchgdetadphi}{\ensuremath {\langle \mathrm{d}^2N_\text{ch}/\mathrm{d}\eta\,\mathrm{d}\phi\rangle}\xspace}
\newcommand{\dpTsumdetadphi}{\ensuremath{\langle \mathrm{d}^2\sum p_t/\mathrm{d}\eta\,\mathrm{d}\phi \rangle}\xspace}
\newcommand{\ptlead}{\ensuremath{p_{\perp}^{\rm lead}}\xspace}

\newcommand{\ptminnought}{\ensuremath{p_{\perp,0}^{\text{min}}}\xspace}

\newcommand{\chisq}{\ensuremath{\chi^2}\xspace}
\newcommand{\wOb}{\ensuremath{w_{\mathcal{O},b}}\xspace}
\newcommand{\p}{\ensuremath{\vec{p}}\xspace}

\newcommand{\preco}{\ensuremath{p_{\rm reco}}\xspace}

\newcommand{\EEiiiiCTEQ}{\textsc{ue-ee-4-cteq6l1}\xspace}
\newcommand{\EEvCTEQ}{\textsc{ue-ee-5-cteq6l1}\xspace}

\title{\boldmath Constraining MPI models using \seff and recent Tevatron and LHC Underlying Event data}

\author[a]{M. H. Seymour}
\author[a]{A. Si\'odmok}

\affiliation[a]{Consortium for Fundamental Physics,
		School of Physics and Astronomy, 
		\\ The University of Manchester, Manchester, M13 9PL, U.K.}

\emailAdd{michael.seymour@manchester.ac.uk}\emailAdd{andrzej.siodmok@manchester.ac.uk}

\abstract{
We review the modelling of multiple interactions in the event generator \herwig{++} 
and study implications of recent tuning efforts to Tevatron and LHC data. 
It is often said that measurements of the effective cross section for
double-parton scattering, \seff, are in contradiction with
models of the final state of multi-parton interactions, but we show that 
the \herwig{++} model is consistent with both and gives stable predictions for 
underlying event observables at 14~\TeV. 
}
\preprint{{\flushright MAN/HEP/2013/14\\ MCnet-13-08\\ }}
\begin{document}

\maketitle
\flushbottom
\section{Introduction}
\label{sec:intro}
In QCD, high momentum-transfer reactions in hadron-hadron collisions are
understood as the consequence of high momentum-transfer scattering or
annihilation-production processes between the partonic consituents of
the hadrons. Although the hadronic structure is complicated and
impossible to predict at present, the factorisation theorems of QCD
predict that, for sufficiently inclusive cross sections, this
complication factorises into universal (process-independent but
hadron-dependent) parton distribution functions, convoluted with
(process-dependent but hadron-independent) partonic cross
sections. Thus, the hard cross section is seen as the consequence of one
parton from each hadron interacting perturbatively.

However, in seeking to understand and predict the final states of
individual hadronic collisions, one is interested in more exclusive
observables, which cannot be described in this purely factorised
way. One talks of the underlying event, the rest of the event that
accompanies the products of a hard collision, or of soft inclusive events,
hadron collisions in which no hard collision occurred. Since the
pioneering work of Sj\"ostrand and van Zijl~\cite{Sjostrand:1987su}, the importance of
multi-parton interactions (MPIs) has been recognised in describing both
underlying events and soft inclusive events in terms of multiple
parton-parton interactions within a single hadron-hadron collision. One
pictures, on the Lorentz-contracted time-scale of a collision, each of
the hadrons to be a frozen disc of partons, with individual partons from
one interacting with the other locally and independently. Although other
approaches have been tried~\cite{Marchesini:1988hj}, MPIs are now firmly
established as the primary source of underlying event activity,
particularly at the high energies achieved by the Tevatron and LHC, and
are the basis of all models in current use for LHC physics.

MPIs contribute to the underlying event and soft inclusive event
activity in two ways. Firstly, additional scatters can produce
additional semi-soft partons throughout the event, directly resulting in
additional jet (sometimes called mini-jet) activity, contributing to the
energy- and hadron-flow both within hard final-state jets and between
jets. Secondly, the colour structure of QCD is such that partonic
scattering is dominated by octet colour-exchange. Hence partonic
scattering, whether hard, semi-soft, or so soft that it does not produce
any observable jet activity, generally involves a colour exchange between the two
hadrons and hence soft hadron production at all rapidities in the
event. The contribution of these two mechanisms means that there is a
considerable interplay between the model parameters that describe the
purely-MPI parts of the model, like the distribution of partons across
the face of the hadron (the ``matter distribution''), and those that
describe the hadronisation of the final state, like the string- or
cluster-model parameters, or parameters that describe colour
deconnections, reconnections and interconnections between the different
coloured systems resulting from different partonic scatters. Tuning one
part of the model without the other becomes effectively meaningless and
a given set of MPI model parameters should only be interpreted in the
context of the other model parameters with which they were
tuned\footnote{MPI models typically use inclusive pdf sets to guide
  their multi-parton distribution function models and the same comment
  applies there: a given set of MPI model parameters should be linked
  with the pdf set with which they were tuned.}.

Of course, to get a deeper understanding of MPIs and to separate the two
aspects of their influence on final states, one would like to make a
more direct measurement of double-parton events in which two (or more)
independent parton-parton scatters are clearly identified. This is made
very difficult by the fact that the double-parton scattering cross
sections are largest for soft jets for which higher-order and
non-perturbative corrections are large and the jet distributions are
only quite weakly correlated with the underlying parton
distributions. The principle behind all pre-LHC measurements is to use
dijet or photon+jet pairs, which, in the MPI model at the partonic
level, produce two back-to-back pairs each independently flat in
azimuth, whereas the background from a single partonic event is strongly
peaked. The more recent ATLAS measurement used W production and one
dijet scatter, while future studies of $\mathrm{W}^+\mathrm{W}^+$ events
promise almost background-free measurements.

The measurements of double-parton scattering are typically phrased in
terms of an effective cross section parameter, \seff, defined as
follows. One measures the cross sections for events that contain
scatters of given types $\mathrm{a}$ or $\mathrm{b}$,
$\sigma_{\mathrm{a}}$ and $\sigma_{\mathrm{b}}$, and of events that
contain scatters of both types $\mathrm{a}$ and $\mathrm{b}$,
$\sigma_{\mathrm{ab}}$. \seff is then defined through\footnote{If
  $\mathrm{a}$ and $\mathrm{b}$ are indistinguishable, as in 4-jet
  production, a statistical factor of $\frac12$ must be inserted.}
\begin{equation}
  \sigma_{\mathrm{ab}} =
  \frac{\sigma_{\mathrm{a}}\,\sigma_{\mathrm{b}}}{\seff}\,.
\end{equation}
In the case that $\sigma_{\mathrm{a,b,ab}}$ are inclusive cross
sections, \seff is independent of $\mathrm{a}$ and $\mathrm{b}$.

A first estimate of \seff was made by the AFS
experiment~\cite{Akesson:1986iv}, followed by an unsuccessful search by
UA2~\cite{Alitti:1991rd} and a first Tevatron measurement by
CDF~\cite{Abe:1993rv},
but the first precise measurements were made by
the Tevatron experiments~\cite{Abe:1997xk,Abazov:2009gc}. First LHC
measurements have been made by 
LHCb~\cite{Aaij:2011yc,Aaij:2012dz} and ATLAS~\cite{Aad:2013bjm}. We review these
results below. Within the experimental uncertainties, there is no
indication that \seff is energy dependent and we quote an
energy-independent average of
\begin{equation}
 \seff = (13.9\pm1.5)~\mbox{mb}.
\end{equation}

Within MPI models, the value of \seff is closely related to the matter
distribution. In fact, having fixed the parameters of a given MPI model,
one can make an unambiguous prediction of~\seff. Within various
different tunes of various different models, one obtains values in the
range 20 to 40~mb. For example:
\begin{itemize}
 \item \herwig{++} tune UE-EE-4-CTEQ6L1~\cite{Gieseke:2011xy,Gieseke:2011na} gives \seff = 25.4~mb,
 \item \herwig{++} tune UE-EE-4~\cite{Gieseke:2011xy,Gieseke:2011na} gives \seff = 30.9~mb,
 \item \herwig{++} tune UE-EE-SCR-CTEQ6L1~\cite{Gieseke:2012ft,Arnold:2012fq} gives \seff = 22.8~mb,
 \item \pythia{ 8} tune 4C~\cite{Corke:2010yf,torbjorn} gives \seff = 33.7~mb, and
 \item Pythia 6 with various tunes (D6T, Z1, Perugia~\cite{Skands:2009zm,paolo}) gives \seff
   values between 20 and 30~mb.
\end{itemize}
These values are all clearly above the experimental value, a fact that
has been used to argue that the Monte Carlo MPI models are
oversimplified and should be improved, for example by including
$x$-dependence~\cite{Strikman:2011cx,Frankfurt:2010ea,Corke:2011yy} or
$p_\perp$-dependence~\cite{Blok:2013bpa}.

In this paper, we explore the interplay between the model parameters by
taking a different approach. Rather than tuning the parameters to the
final-state data, fixing them at their best-fit values and comparing the
prediction of \seff with data, we instead include the value of \seff
and its experimental uncertainty as one of the pieces of data that we
fit to. Within the fitting framework we can then explore whether there
is a tension between \seff and the final-state data and the extent to
which we can find a set of model parameters that describe both well.

This approach is mandated by the ``Summary of the Workshop on
Multi-Parton Interactions (MPI) 2012''~\cite{Abramowicz:2013iva} whose
``to do list'' for MPI and Monte Carlo development starts with ``The
value of \seff should serve as a constraint on the Monte Carlo models
since the recent tunes of MPI models to the LHC data predict its value
to be between 25--42~mb''.

We will also study the predictions of our fits for the extrapolation to
14~TeV. The remainder of this paper is set out as follows. In
Sect.~\ref{sec:model}, we will set out the particular variant of MPI
model that we use, as implemented in \herwig{++}. In
Sect.~\ref{sec:constraining} we briefly review the experimental
measurements of \seff and the constraint they impose on the overlap
function in our model. In Sect.~\ref{sec:tuning} we discuss the tuning
of our model parameters to the value of \seff and the underlying event
data from CDF and ATLAS and show the extrapolation of the tunes to
14~\TeV. In Sect.~\ref{sec:conclusions} we make some concluding remarks
and in Appendix~\ref{Appendix} we show and discuss some more detailed
technical aspects of our tunes.

\section{MPI model}
\label{sec:model}
The MPI model used in
\herwig{++}~\cite{Butterworth:1996zw,Borozan:2002fk,Bahr:2008dy,Bahr:2008pv}
has been reviewed several times. Our intention here is not to review it
again, but just to provide enough background to set our discussion of
the tuning of its parameters in context.

The model is formulated in impact parameter space. At fixed impact
parameter, multiple parton scatterings are assumed to be independent.
Parton-parton scatterings are divided into soft and hard by a parameter
\ptmin, which is one of the main tuning parameters in our model. Above
\ptmin, scatters are assumed to be perturbative, and take place
according to leading order QCD matrix elements convoluted with inclusive
pdfs and a matter distribution~$A(b)$,
\begin{equation}
  A(b) = \int \mathrm{d}^2b_1 \, G(b_1) \int \mathrm{d}^2b_2 \, G(b_2)
  \, \delta^2(\mathbf{b}-\mathbf{b}_1+\mathbf{b}_2) \, ,
\end{equation}
where
\begin{equation}
  G(b) = \frac{\mu^2}{4\pi}(\mu b)K_1(\mu b)
\end{equation}
is the Fourier transform of the electromagnetic form factor, giving
\begin{equation}
  A(b) = \frac{\mu^2}{96\pi}(\mu b)^3K_3(\mu b),
\end{equation}
with $K_i(x)$ the modified Bessel function of the $i$th kind. The
parameter $\mu^2$ appearing in $G(b)$ is another of the main tuning
parameters and plays the role of an effective inverse proton
radius. Note that $G(b)$, and therefore also $A(b)$, is normalised to
unity. We allow the value of
\ptmin to vary with energy according to
\begin{equation}
  \ptmin(s) = \ptminnought \left(\frac{\sqrt{s}}{E_0}\right)^b,
\label{eq:power}
\end{equation}
and, in fact, it is $\ptminnought$ and $b$ that we fit to data, with
$E_0=7$~\TeV.

Below \ptmin, scatters are assumed to be non-perturbative, with
``Gaussian'' transverse momentum distribution and valence-like
longitudinal momentum distribution. The matter distribution is assumed
to have the same form as above, but we allow it to have a different
$\mu^2$ value. We therefore label it as $A_{\textit{soft}}(b)$. We call
the transverse momentum distribution ``Gaussian'', because once the
parameters are fitted to data, it turns out that the width of the
Gaussian is imaginary and hence the distribution in transverse momentum
is actually peaked at \ptmin and small at very small $p_{\perp}$.

The inelastic hadron-hadron cross section serves as a constraint on the
parameters, given by
\begin{equation}
  \sigma_{\textit{inelastic}} = \int \mathrm{d}^2b
  \left(1-\mathrm{e}^{-A(b)\sigma_{\textit{hard}}
    -A_{\textit{soft}}(b)\sigma_{\textit{soft}}}\right),
\end{equation}
as does the inelastic slope parameter. The differential parton-parton
cross section is required to be continuous at \ptmin. Thus, given the
parameters \ptmin and $\mu^2$, the values of $\mu^2_{\textit{soft}}$,
$\sigma_{\textit{soft}}$ and its Gaussian width are fixed.

The probability distribution of number of scatters is Poissonian at a
given value of impact parameter, but the distribution over impact
parameter gives a considerably longer than Poissonian tail. The number
of soft and hard scatters is chosen according to this distribution and
generated according to their respective distributions. Each hard scatter
is evolved back to the incoming hadron according to the standard parton
shower machinery. Energy-momentum conservation is imposed at this stage
by rejecting any scatters that take the total energy extracted from the
hadron above its total energy. The individual scatters are either colour
connected with each other in a random (i.e.\ uncorrelated with their
hardness) sequence, or disconnected from other scatters, with
probability \pdisrupt. A~colour reconnection model,
described in detail in Ref.~\cite{Gieseke:2012ft}, with reconnection
probability \preco is applied. The parameters we tune to data are
therefore \ptminnought\ and $b$ from Eq.~(\ref{eq:power}), $\mu^2$,
\pdisrupt and \preco.

Within this model, the probability of $n$ perturbative scatters of type
$\mathrm{a}$ and $m$ of type $\mathrm{b}$ is
\begin{equation}
  \sigma_{n\mathrm{a},m\mathrm{b}} = \int \mathrm{d}^2b\,
  \frac{\left(\sigma_\mathrm{a}A(b)\right)^n}{n!}\,
  \frac{\left(\sigma_\mathrm{b}A(b)\right)^m}{m!}\,
  \mathrm{e}^{-A(b)(\sigma_\mathrm{a}+\sigma_\mathrm{b})}.
\end{equation}
The inclusive $\mathrm{ab}$ cross section is then
\begin{equation}
  \sigma_{\mathrm{ab}} = \int \mathrm{d}^2b
  \left(\sigma_\mathrm{a}A(b)\right)
  \left(\sigma_\mathrm{b}A(b)\right),
\end{equation}
so that
\begin{equation}
  \seff = \frac1{\int \mathrm{d}^2b\,A(b)^2} = \frac{28\pi}{\mu^2}\,.
  \label{eq:seffvalue}
\end{equation}
That is, within our model, a measurement of \seff translates directly
into a measurement of~$\mu^2$.

\section{Measurements of \seff}
\label{sec:constraining}
In this section we briefly review the experimental measurements of \seff
made to date.

The first evidence for double parton scattering was obtained by the AFS
collaboration \cite{Akesson:1986iv} from 4-jet events in $pp$ collisions
at $\sqrt{s}=\unit{63}{\GeV}$. They extracted a value of
$\seff=\unit{5}{\mbox{mb}}$, without quoting an uncertainty. It is
interesting to
note that they stated their expectation as $\seff=\unit{13}{\mbox{mb}}$.

The UA2 collaboration performed a search for double parton scattering
\cite{Alitti:1991rd} in 4-jet events in $p\bar{p}$ collisions at
$\sqrt{s}=\unit{630}{\GeV}$ but only set a limit,
$\seff>\unit{8.3}{\mbox{mb}}$ at
95\% confidence level. They stated their expectation as
$\seff<\unit{40}{\mbox{mb}}$.

The CDF collaboration made a first measurement from 4-jet events in
$p\bar{p}$ collisions at $\sqrt{s}=\unit{1.8}{\TeV}$\cite{Abe:1993rv}
quoting a value of $\seff=12.1^{+10.7}_{-\phantom{1}5.4}\,\mathrm{\mbox{mb}}$
(stating their expectation as $\seff=\unit{22}{\mbox{mb}}$).

\texttt{\begin{figure*}[]
\begin{center}
 \includegraphics[width=0.65\textwidth]{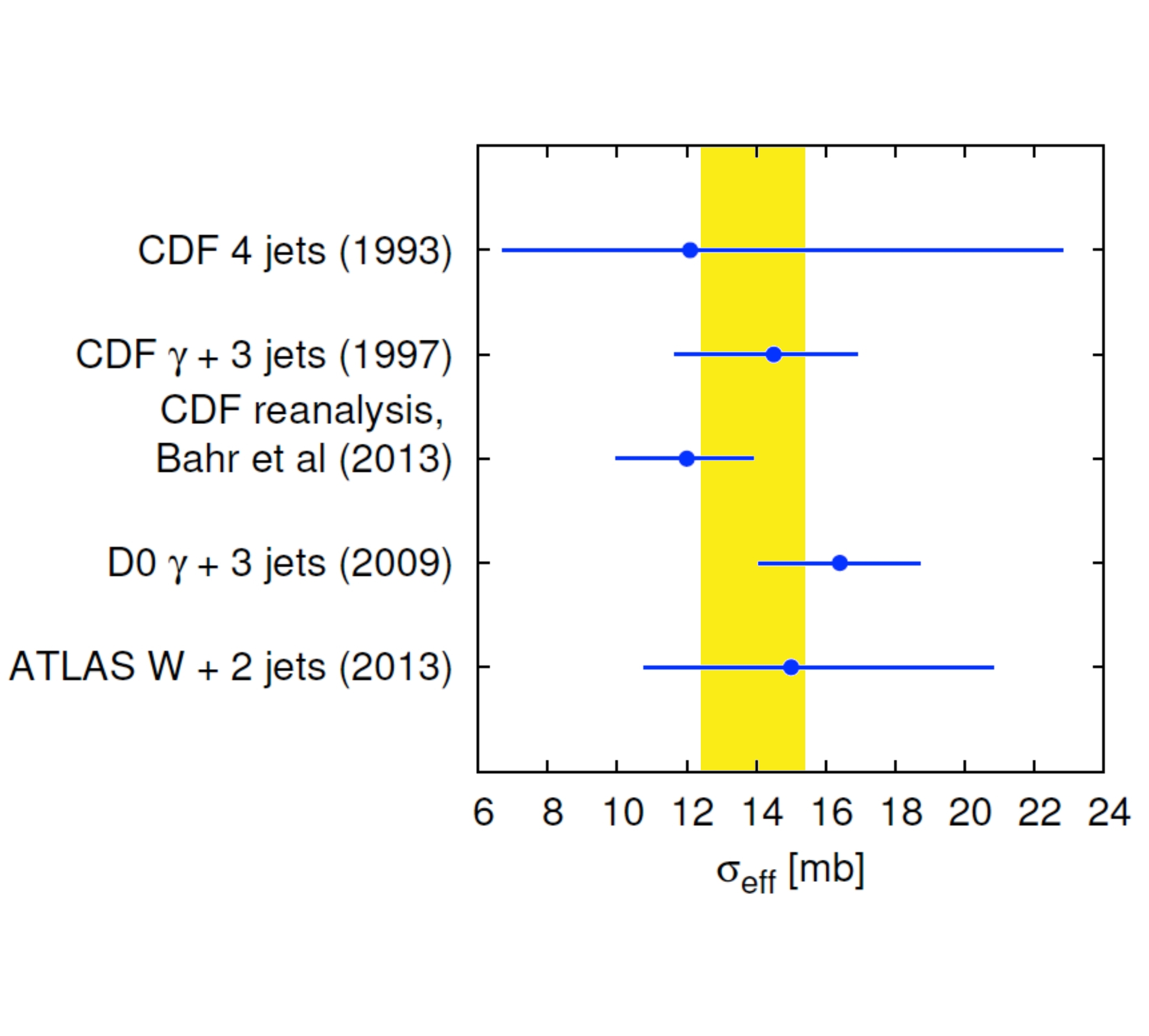}
 \vspace*{-10ex}
 \caption{Measurements of \seff, as described in the text. The yellow
   band shows the weighted average of D0 and reanalysed CDF results.}
  \label{fig:MeasurementsSigma}
\end{center}
\end{figure*}}
Four years later, CDF were able to make a much more precise measurement,
from $\gamma+3$-jet events~\cite{Abe:1997xk}. They extracted a value of
\begin{equation}
 \seffcdf = (14.5\pm1.7^{+1.7}_{-2.3})~\mbox{\mbox{mb}},
\end{equation}
with their expectation stated as $\seff=\unit{11}{\mbox{mb}}$.
However, as first pointed out by Treleani~\cite{Treleani:2007gi}, this
CDF measurement used a non-standard definition of the double-parton
scattering cross section that makes the value of \seff dependent on the
individual scattering cross sections. Based on a theoretically-pure 
(parton-level) analysis, Treleani estimated the inclusive (process
independent) value as
\begin{equation}
 \seff=10.3~\mbox{\mbox{mb}}.
\end{equation}
This result would be correct under the assumption that CDF were able to uniquely identify 
and count the nu\mbox{mb}er of scatters in an event, which is certainly not the case.
In our recent paper, Ref.~\cite{Bahr:2013gkj}, we re-analyzed CDF's event
definition to provide an improved correction leading to
\begin{equation}
 \seffcdfre = (12.0 \pm 1.4 ^{+1.3}_{-1.5})~\mbox{\mbox{mb}}.
\end{equation}

The D0 collaboration performed a similar analysis of $\gamma+3$-jet
events in $p\bar{p}$ collisions~\cite{Abazov:2009gc} and were able to
extract independent \seff values in each of three $p_{\perp}$ bins,
yielding an average value of
\begin{equation}
 \seffd = (16.4\pm0.3\pm2.3)~\mbox{\mbox{mb}},
\end{equation}
and was the first paper not to quote an expectation.

A first measurement has been made at the LHC by the ATLAS experiment
using $W+2$-jet production in $pp$ collisions at
$\sqrt{s}=\unit{7}{\TeV}$, yielding a value of
\begin{equation}
 \sigma_{\textit{eff,ATLAS}} = (15\pm3^{+5}_{-3})~\mbox{\mbox{mb}},
\end{equation}
Although not yet as precise as the Tevatron measurements, it lends
weight to the assumption that \seff is energy independent. There is
clear promise for future more precise measurements.
\texttt{\begin{figure*}[htb]
  \includegraphics[width=0.49\textwidth]{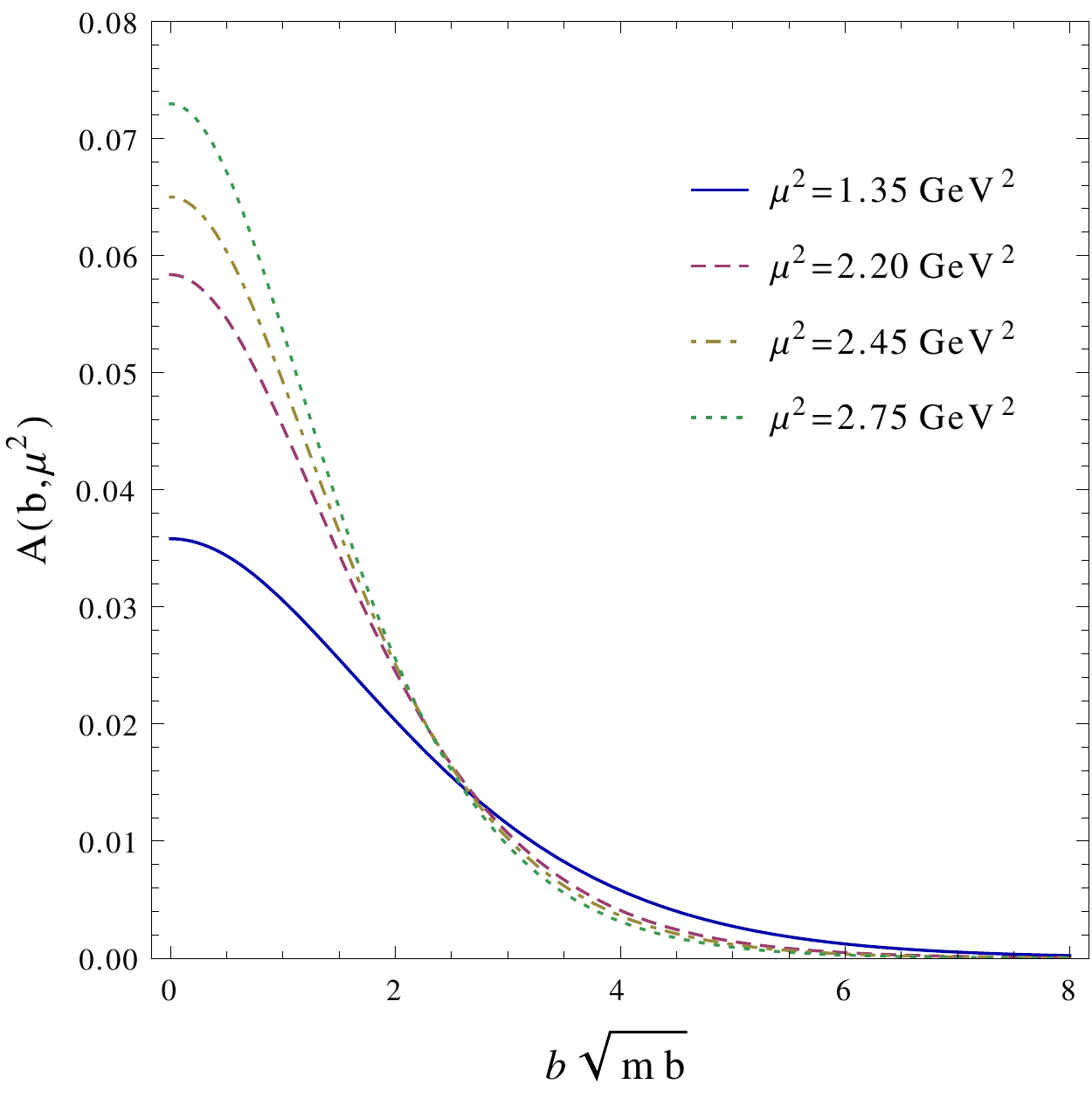}\hfill
  \includegraphics[width=0.5\textwidth]{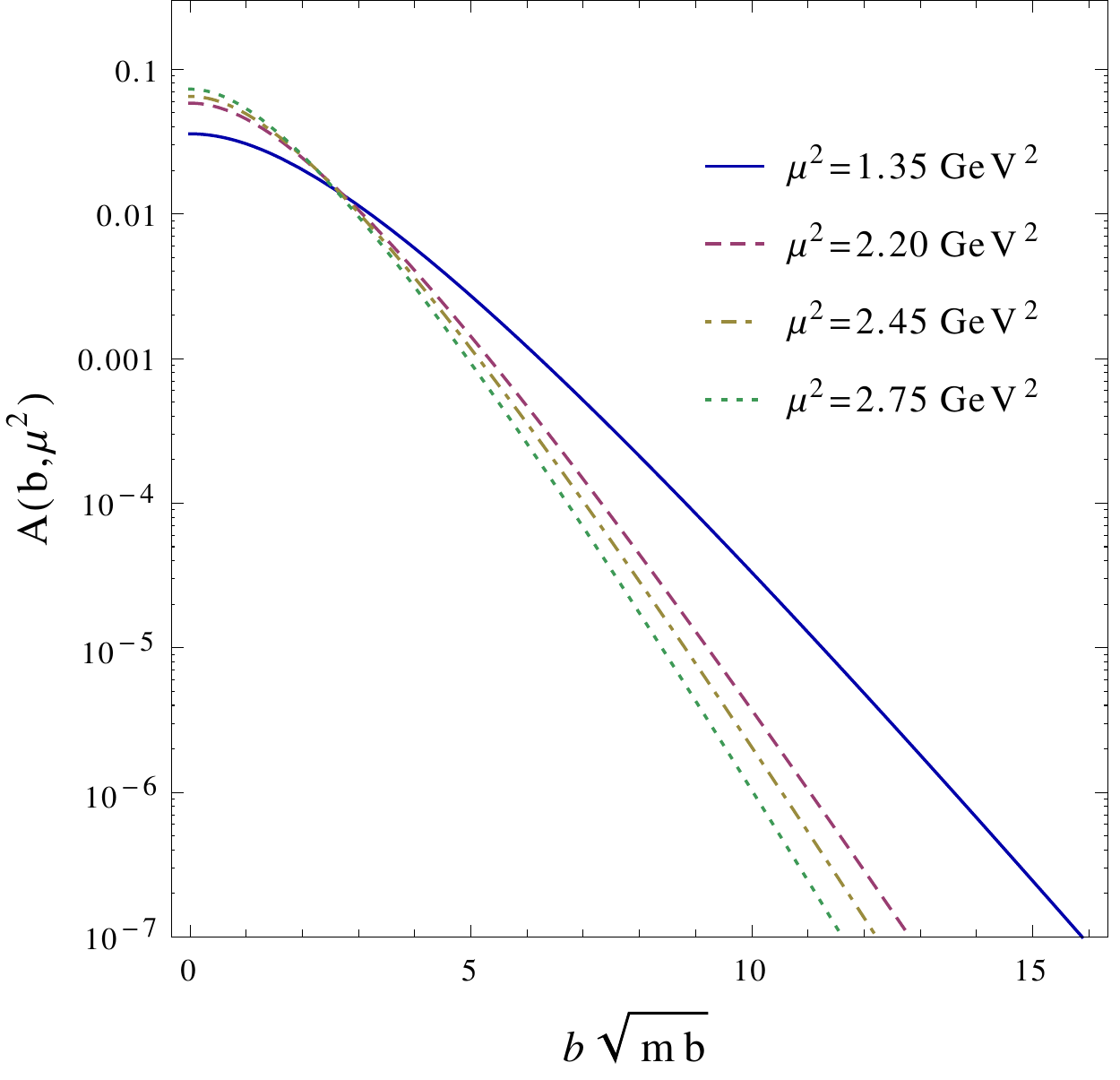}
  \caption{The overlap function $A(b)$ for the central value of $\mu^2$
  given by \seffdcdf, together with its $1\sigma$ upper and lower
  bounds, and from the previous tune \EEiiiiCTEQ
  ($\mu^2=1.35~\mbox{GeV}^2$).}
  \label{fig:Overlap}
\end{figure*}}

The first measurement of double-charm production was made by the LHCb
collaboration in \cite{Aaij:2011yc} and followed up by measurements in
many different charm channels~\cite{Aaij:2012dz}. The results, summarised
in Fig.~10 of Ref.~\cite{Aaij:2012dz}, are that the individual channels
have uncertainties that are, at present, significantly larger than those
from the Tevatron experiments, that the channels involving a J/$\psi$
are in agreement with the Tevatron measurements, but that the channels
involving two open charm quarks are much higher and most of those
involving an open charm quark and an open anticharm quark are much
lower. In Ref.~\cite{Maciula:2013kd}, this discrepancy was attributed to
$k_\perp$-factorization effects, while in Ref.~\cite{Berezhnoy:2012xq}
it was claimed that there is no discrepancy when experimental acceptance
is properly taken into account. We do not consider these results to be
understood at present, so we do not include them in our analysis.

In order to give a co\mbox{mb}ined result, we take the weighted average of the
two most precise results, the D0 and reanalysed CDF ones:
\begin{equation}
 \seffdcdf = (13.9\pm1.5)~\mbox{\mbox{mb}}.
\end{equation}
The agreement of this co\mbox{mb}ined result with the independent measurements
is displayed in Fig.~\ref{fig:MeasurementsSigma}.

In our form factor model, this corresponds to 
\begin{equation}
\label{eq:mu2}
 \mu^2 \sim (2.45^{+0.30}_{-0.25})~\mbox{GeV}^2.
\end{equation} 
The shapes of the overlap function for those values of $\mu^2$ are shown
in Fig~\ref{fig:Overlap}. The value from the previous tune of
\herwig{++}, \EEiiiiCTEQ, with $\mu^2=\unit{1.35}{\GeV^2}$, is shown for
comparison. It is clearly well outside the range given by the \seff
measurements.

\section{Tuning to the underlying event data}
\label{sec:tuning}
In this section we address the question of whether the MPI model in \herwig{++} 
is consistent with the \seff and underlying event (UE) data. To that end we need to check whether 
it is possible to find values of the free parameters of the MPI model 
that allow a good description of the experimental data. As described in
Sect.~\ref{sec:model}, the MPI model in \herwig{++} has five parameters
that we consider freely tunable: 
$\pdisrupt$, $\preco$, $\ptminnought$, $b$ and~$\mu^2$.
Because we are dealing with a large number ($N=5$) of tunable
parameters, a simple tuning strategy of subdividing the $N$-dimensional
parameter space into a grid and for each of the parameter points on this grid 
running \herwig{++} and calculating the total $\chi^2$ against the
experimental data, is ineffective.
A comprehensive scan of 5 parameters, with 10 divisions in each
parameter, would require too much CPU time.

Instead, we use the parametrization-based tune method provided by the
\Professor package~\cite{Buckley:2009bj}, which is much more efficient
for our case.  The starting point for this tuning procedure is the selection of
a range $[p_i^{\rm min},\, p_i^{\rm max}]$ for each of the $N$ tuning
parameters~$p_i$. Event samples are generated for random points in this
$N$-dimensional hypercube in parameter space.  The number of points
sampled is chosen, depending on the number of input parameters, to
ensure good control of the final tune. Each generated event is directly
handed over to the Rivet package~\cite{Buckley:2010ar}, which implements
the experimental analyses.  Thus the results for each observable are
calculated at each set of parameter values. \Professor parametrizes each
bin of each histogram as a function of the input parameters. It is then
able to find the set of parameters that fits the selected observables
best. As a user, one simply has to choose the set of observables that
one wishes to tune to and, optionally, their relative weights in the fit.

\subsection{Observables}
On top of the \seff data mentioned above we examine a set of standard UE
observables at different collider energies whose description is sensitive to
the MPI model parameters. The standard UE measurements are made relative to a leading 
object (the hardest charged track or jet). Then, the transverse plane is subdivided in 
azimuthal angle $\phi$ relative to this leading object at $\phi=0$.  The region around 
the leading object, $|\phi|<\pi/3$, is called the ``towards'' region.  
The opposite region, where we usually find a recoiling hard object, $|\phi| > 2\pi/3$, 
is called the ``away'' region, while the remaining region, transverse to
the leading object and its recoil, called the ``transverse'' region, is
expected to be the most sensitive to MPI
activity. The UE experimental data used for the tune should be measured
at a wide range of collider
energies in similar phase-space regions and under not too different trigger
conditions. These conditions are met by two UE observables:
\begin{itemize}
 \item The mean number of stable charged particles per unit of
   $\eta$-$\phi$: \dNchgdetadphi, in the tranverse region;
 \item The mean scalar $p_{\perp}$ sum of stable particles:
   \dpTsumdetadphi, in the tranverse region.
\end{itemize}
Both are measured as a function of $\ptlead$ (with $\ptlead <
\unit{20}{\GeV}$) by ATLAS at \unit{900}{\GeV} and \unit{7}{\TeV}
(with $p_{\perp} > \unit{500}{\MeV}$\footnote{For data collected at
  \unit{7}{\TeV} we also used observables with $p_{\perp} >
  \unit{100}{\MeV}$.})\cite{Aad:2010fh} and by CDF at \unit{300}{\GeV}, \unit{900}{\GeV}
and \unit{1960}{\GeV} (with $p_{\perp} > \unit{500}{\MeV}$)\cite{CDFUEscan}.
In both ATLAS and CDF UE analyses, $\ptlead$ denotes the transverse momentum 
of the hardest track\footnote{CMS have also made measurements of the
  underlying event, but using track-based
  jets\cite{Chatrchyan:2011id}. We confine ourselves to ATLAS's
  hardest-track analysis here, since this matches most closely CDF's
  energy scan analyses, giving us the longest lever arm in energy for a
  single analysis type.}.

Between all of the above data, we have a total of 132~histogram bins for
each of $\dNchgdetadphi$ and $\dpTsumdetadphi$ at various collider
energies and $\ptlead$ values to fit the model parameters to, as well
as the value of \seff. By default, \Professor would treat each of these
265~pieces of data equally, and would not pay any more attention to
reproducing the value of \seff than it would to any individual histogram
bin. On the contrary, we would like to treat each \emph{type} of data,
\dNchgdetadphi, \dpTsumdetadphi and \seff, on an equal footing. We
therefore use the option provided by \Professor\footnote{We also
  modified \Professor to use the exact analytical formula for \seff,
  Eq.~(\ref{eq:seffvalue}), rather than the value calculated by the
  Monte Carlo algorithm, which has (small) numerical errors.} to apply
different weights to different observables (see
Eq.~(\ref{eq:chi2})). We give each of the histogram bins unit weight,
while we give \seff a weight of~132~\footnote{We show the
  weight-dependence of our result in Appendix~\ref{app:weight}, where it
  can be seen that any weight value above about~100 would give similar
  results, within 1~$\sigma$ of the input \seff value.}.

As we discussed in the introduction, MPI model parameters should be used
together with the pdf set with which they have been tuned. Thus, in the
\herwig convention, the name of the tune contains the name of the pdf
set. We have used leading order version 6.1 pdfs from the CTEQ
collaboration~\cite{Stump:2003yu} and therefore, to fit the naming
conventions of \herwig, call our final tune ``\EEvCTEQ''\footnote{We
  have also produced a corresponding set using the MSTW LO*
  pdfs~\cite{Martin:2009iq}. Although the tuned parameters differ
  somewhat, the results for \seff and UE observables do not differ
  significantly, so we do not show the results here.}.
In order to study how fitting the value of \seff influences the result
of the tune we also performed a tune without taking into account the
\seff data. We will refer to this tune as ``\textit{No \seff in fit}''.

\subsection{Results}
We begin by showing the weighted $\chi^2$ value calculated by \Professor:
\begin{equation}
  \chisq(\p) = 
  \sum_{\mathcal{O}}  \sum_{b \, \in \, \mathcal{O}}\wOb 
  \frac{ (f^{(b)}(\p) - \mathcal{R}_b)^2 }{ \Delta^2_b }\,,
  \label{eq:chi2}
\end{equation}
where \wOb is the weight for bin $b$ of observable $\mathcal{O}$,
$f^{(b)}(\p)$ is a function of the tuning parameters $\p$, which
parametrizes the Monte Carlo results, $\mathcal{R}_b$ is the reference
(experimental) value for bin $b$ and the error $\Delta_b$ is the total 
uncertainty\footnote{In practice we attempt to generate sufficient events 
at each sampled parameter point that the statistical Monte Carlo error
is much smaller than the reference error for all bins.} for bin~$b$.
Recall that we use $\wOb=1$ except for \seff, which has $\wOb=132$.

\begin{figure*}[htb]
  \vspace{-7ex}
  \includegraphics[width=0.49\textwidth]{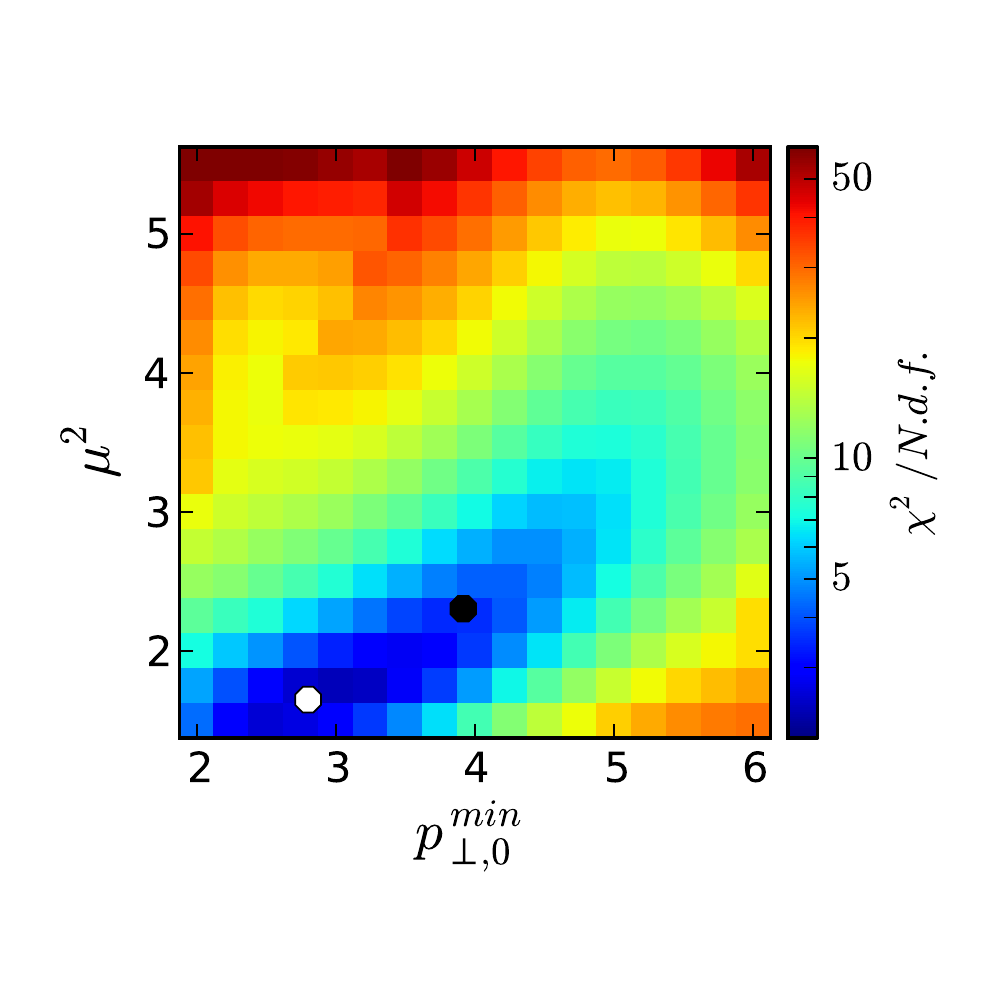}
  \includegraphics[width=0.49\textwidth]{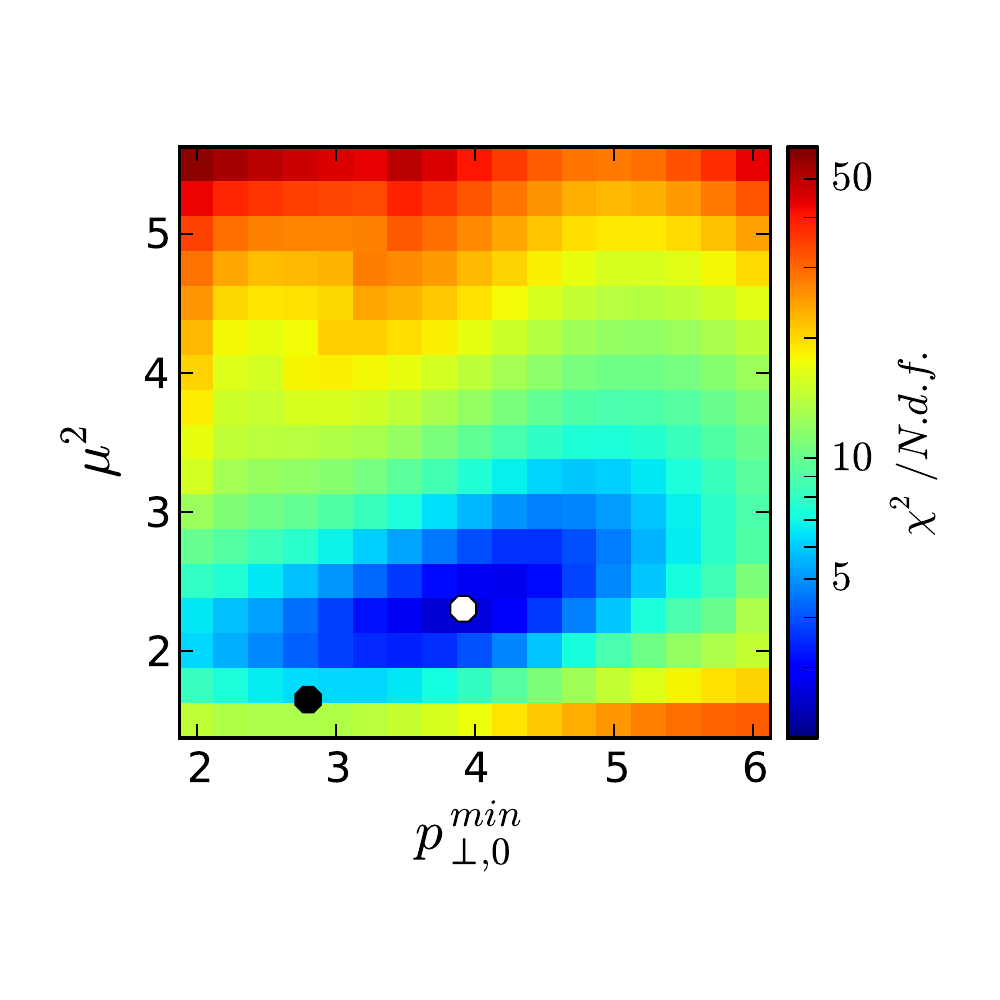}
  \vspace{-6ex}
  \caption{The weighted $\chi^2/N\!.d.f.$ as a function of $\mu^2$ and
    $\ptminnought$. On the left obtained using only the UE data sets,
    on the right taking into account also the \seff data, with weight~132.
    In each plot, its best fit point is shown with a white dot, while
    the best fit point of the other is shown with a black dot for
    comparison. Note that \Professor's definition of ``$N\!.d.f.$''
    includes the weight factors, so the absolute value of $\chi^2$ is
    1.5~times larger in the right plot than in the left, for the same
    colour.}
  \vspace{-2ex}
  \label{fig:chi2}
\end{figure*}

In the left panel of Fig.~\ref{fig:chi2}, we plot the $\chi^2/N\!.d.f.$ value as
a function of $\ptminnought$ and $\mu^2$ fitting only to the UE data,
i.e.~the ``\textit{No \seff in fit}'' fit.
For each parameter pair, we show the minimum $\chi^2/N\!.d.f.$ that can be found
in the corresponding subspace of the other parameters, $\pdisrupt$,
$\preco$ and~$b$.
We see that the $\chi^2/N\!.d.f.$ function forms a long thin valley, with a
strong correlation between $\ptminnought$ and~$\mu^2$. This
reflects the fact that a smaller hadron radius means more multiple
scattering, which can be compensated to give a similar amount of
underlying-event activity by having fewer perturbative MPIs, i.e.~a
larger value of $\ptmin$.
The best fit value is $\ptminnought=\unit{2.80}{\GeV}$,
$\mu^2=\unit{1.65}{\GeV^2}$, but one can obtain good fits up to at least
$\ptminnought=\unit{4.5}{\GeV}$, provided $\mu^2$ is adjusted
accordingly.

In the right panel of Fig.~\ref{fig:chi2}, we plot the equivalent
results including the \seff measurement with a weight of~132,
i.e.~the $\EEvCTEQ$ fit.
As would be expected, the inclusion of the \seff data helps to break the
degeneracy between $\ptminnought$ and $\mu^2$, favouring the $\mu^2$
range that reproduces the \seff value. On the other hand the description
of the UE data is not significantly worsened, as can be seen from the
fact that the best fit with \seff lies `on the valley floor' of the fit
without \seff.

It is worth making several comments here on the values of
$\chi^2/N.d.f.$ obtained in these plots. Firstly, \Professor's
definition of ``$N.d.f$'' includes the weight factors and, hence, is 1.5
times larger in the plot on the right than the plot on the
left. Secondly, while the experimental data points have moderate
point-to-point correlations, the theoretical predictions are strongly
correlated: the shapes of the curves are reasonably stable predictions
and it is mainly their normalization that varies in response to the
variation of parameters. Our aim is to get a reasonable description of
that normalization across a wide range of collider energies. We
therefore can definitely not interpret $\chi^2$ in a strict statistical
sense. Finally, the value plotted is \Professor's prediction of the
$\chi^2/N.d.f.$ at a given point, based on a parametrization of the
Monte Carlo results, which have statistical errors. The $\chi^2/N.d.f.$
values themselves therefore have errors. For example, in
Appendix~\ref{app:stability} we compare a Monte Carlo run at the best
fit point (with $\chi^2/N.d.f.=3.73$) with the Professor prediction for
it (with $\chi^2/N.d.f.=3.51$). We consider anywhere in the valley floor
with $\chi^2/N.d.f.$ up to about 5 in the left-hand plot to be a
reasonable description of the underlying event data.
\begin{figure*}[htb]
  \includegraphics[width=0.5\textwidth]{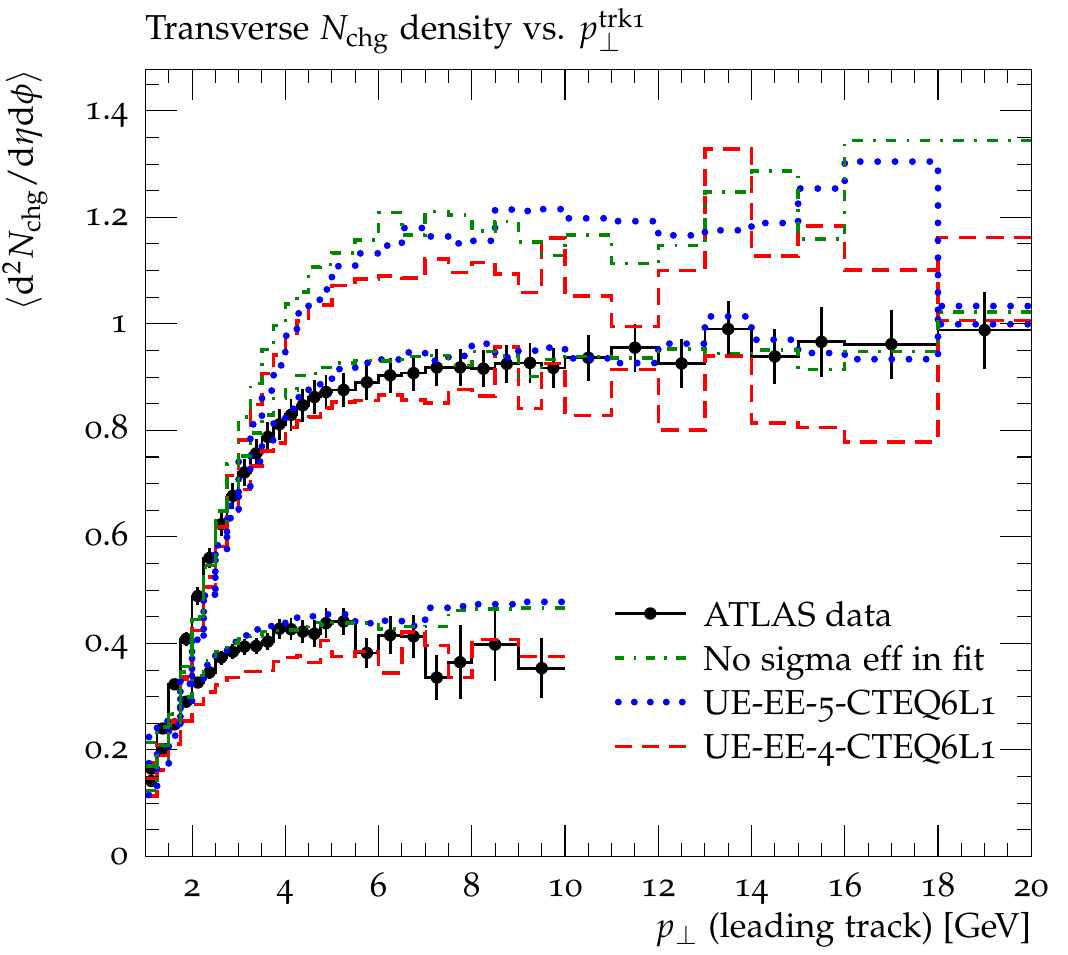}
  \includegraphics[width=0.5\textwidth]{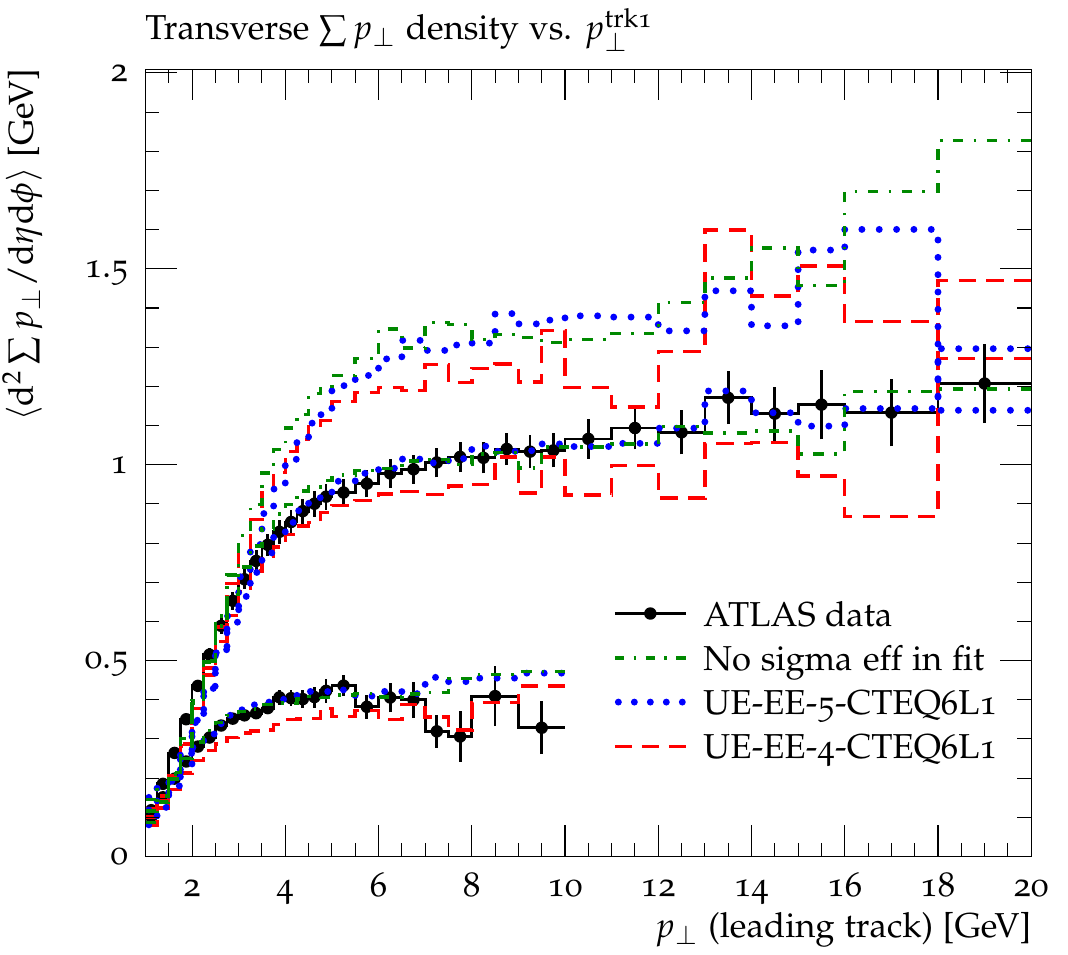}
  \vspace*{-5ex}
    \caption{ATLAS data and model fits at ${900}$~\GeV\ (lowest sets)
      and ${7}$~\TeV\ (middle sets), showing the mean multiplicity
      density (left panel) and mean scalar $p_{\perp}$ sum (right panel)
      of the stable charged particles in the ``transverse'' area as a
      function of~$\ptlead$. The predictions for 14~\TeV\ (top sets) are
      shown for comparison.}
  \label{fig:LHC}
\end{figure*}
\begin{figure*}[htb]
  \includegraphics[width=0.5\textwidth]{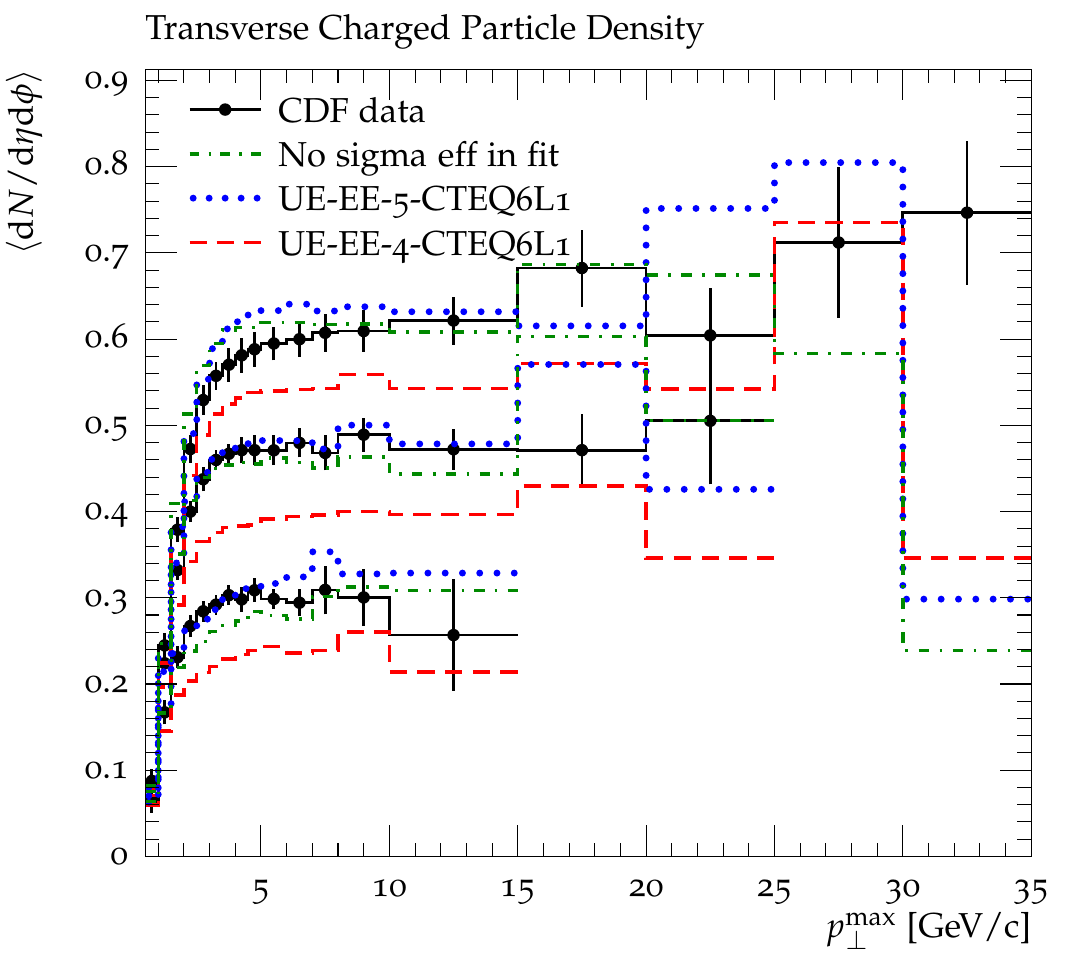}
  \includegraphics[width=0.5\textwidth]{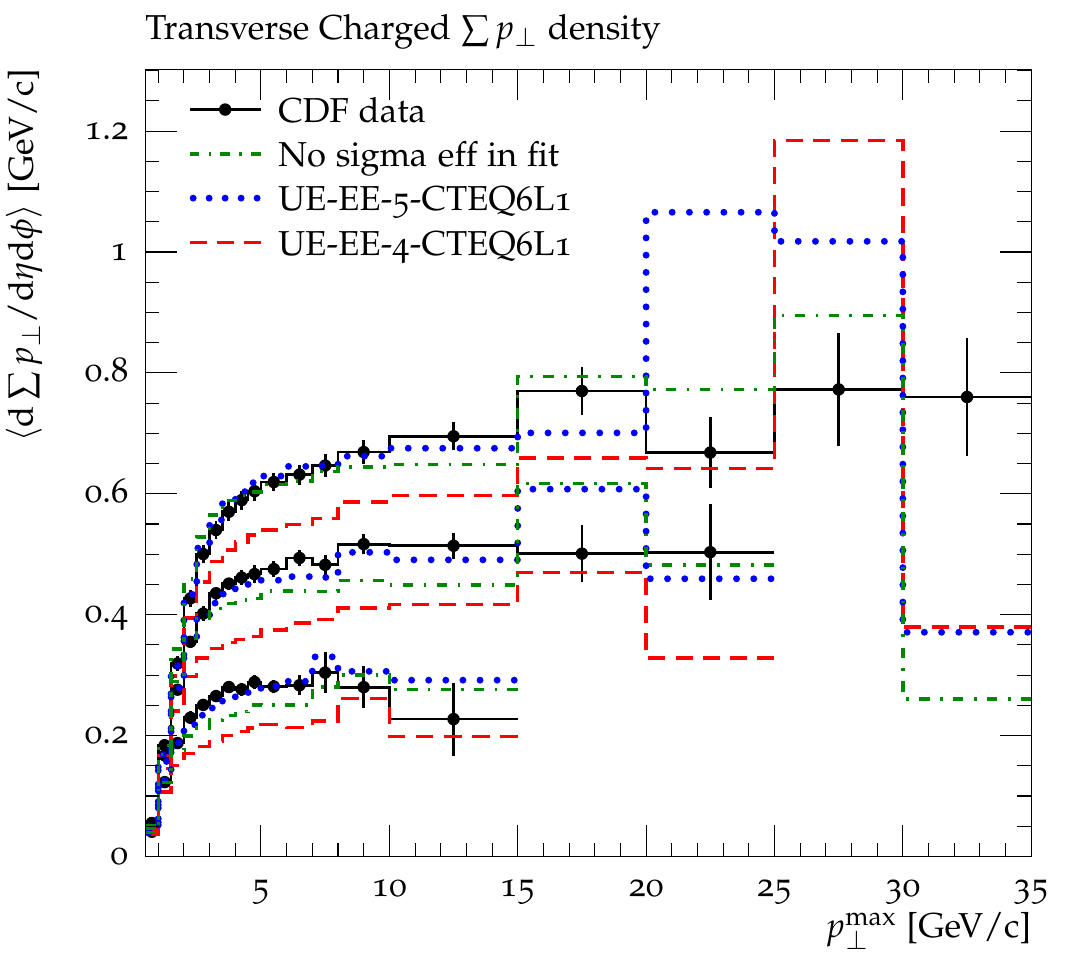}
  \vspace*{-5ex}
    \caption{CDF data and model fits at at ${300}$~\GeV\ (lowest sets),
      ${900}$~\GeV\ (middle sets) and ${1960}$~\GeV\ (top sets), showing
      the mean multiplicity density (left panel) and mean scalar
      $p_{\perp}$ sum (right panel) of the stable charged particles in
      the ``transverse'' area as a function of~$\ptlead$.}
  \label{fig:TVT}
\end{figure*}

In Figs.~\ref{fig:LHC} and~\ref{fig:TVT}, we show the description of the
UE data from ATLAS and CDF respectively. We see that the ``\textit{No
  \seff in fit}'' and $\EEvCTEQ$ fits do indeed describe the data equally
well, and significantly better than the previous $\EEiiiiCTEQ$ fit,
which is the default tune of \herwig{++}~2.6~\cite{Arnold:2012fq}.
The values of the tuned parameters are given in Table~\ref{tab:fixedparams}.

Figure~\ref{fig:LHC} also shows the prediction for results at
14~\TeV\ where, again, the results of the ``\textit{No \seff in fit}'' and
$\EEvCTEQ$ fits are very similar, showing that requiring the fits to
describe \seff has not biased the UE results significantly.

\begin{table*}[htb]
\begin{center}
  \begin{tabular}{lccc} \toprule & \EEiiiiCTEQ & ``\textit{No \seff in fit}'' & \EEvCTEQ \\
    \midrule
    $\mu^2\enspace [\GeV^2]$      &   1.35       & 1.65  &  2.30    \\
    $\pdisrupt$                   &   0.75       & 0.22  &  0.80    \\
    $\preco$                      &   0.61       & 0.60  &  0.49    \\
    \midrule
    $\ptminnought\enspace [\GeV]$ &   2.81       & 2.80  &  3.91    \\
    $b$                           &   0.24       & 0.29  &  0.33    \\
    \bottomrule
  \end{tabular}
  \caption{Parameters of the underlying event tunes. The
  last two parameters describe the running of $\ptmin$ according to
  Eq.~(\ref{eq:power}).}
  \label{tab:fixedparams}
 \end{center}
\end{table*}

\begin{figure*}[htb]
\begin{center}
  \vspace*{-5ex}
  \includegraphics[width=0.9\textwidth]{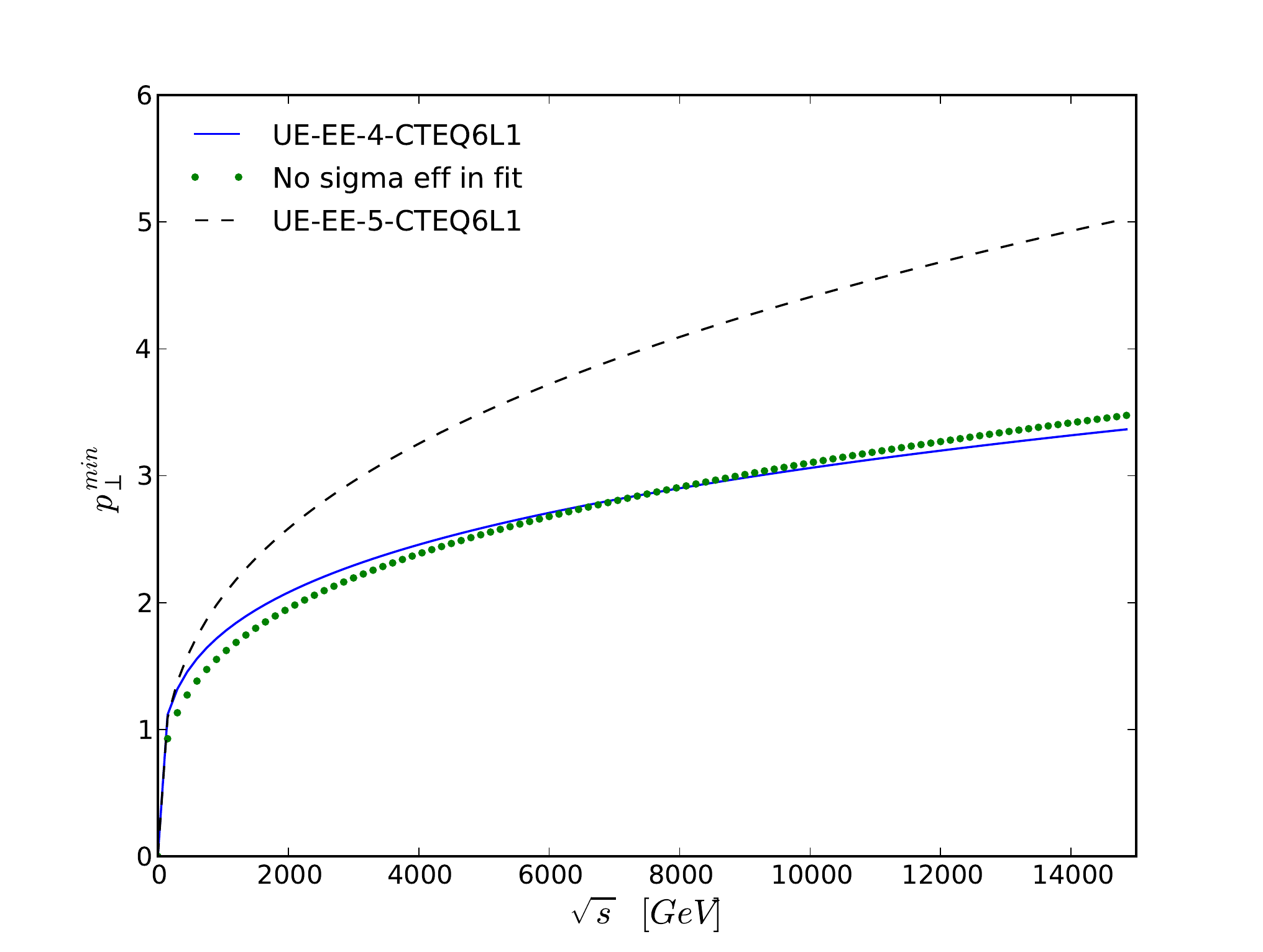}
  \vspace*{-2ex}
  \caption{Collision energy dependence of $\ptmin$ in the three fits
    discussed in the text.}
  \vspace*{-5ex}
  \label{fig:ptmin}
\end{center}
\end{figure*}

Finally, in Fig.~\ref{fig:ptmin}, we show the dependence of $\ptmin$ on
the collision energy. The ``\textit{No \seff in fit}'' fit is very
similar to the previous default, whereas we have seen that their UE
predictions are quite different, owing to the different values of the
other parameters. On the other hand, the ``\textit{No \seff in fit}''
fit and our new best fit, \EEvCTEQ, are quite different, but the
correlation with the other fit parameters means that their UE
predictions are very similar. This clearly shows the importance of the
simultaneous fit of all parameters and the danger of ascribing too much
physical significance to the value of one parameter in isolation of the
others.

\section{Conclusions}
\label{sec:conclusions}
We have discussed the experimental measurements of \seff from
double-parton scattering. We concluded that, over the energy range from
the Tevatron to the LHC, there is no evidence that \seff is energy
dependent. The best value of \seff at present comes from a combination
of the CDF and D0 measurements.

We have also studied fits of our MPI model parameters to underlying
event data from CDF and ATLAS over the collision energy range from
300~\GeV\ to 7~\TeV. It is commonly stated that there is a conflict
between such fits and the value of \seff. Indeed, we have seen that our
best fit to the UE data yields $\seff=\unit{20.6}{\mbox{mb}}$, almost 5$\sigma$
from its measured value.

However, we have also seen that by including \seff in the fit, with the
same weight as the sum of each of the multiplicity and $p_\perp$ sum
data, we are able to obtain a set of parameters that yields
$\seff=\unit{14.8}{\mbox{mb}}$, less than 1$\sigma$ from its measured value,
without significantly worsening the description of the underlying event
data. The extrapolation to 14~\TeV\ is likewise unaffected. It is worth
noting that other model parameters are significantly different between
the two tunes, emphasising the importance of simultaneous tuning.

We conclude that it is possible to describe the double-parton scattering
cross section and underlying event final states with a common set of
parameters in the \herwig MPI model. The parameter set \EEvCTEQ\ will be
the default set with the next \herwig{++} release.

\vspace{4ex}
\acknowledgments
We are grateful to the Cloud Computing for Science and Economy project (CC1) 
at IFJ PAN (POIG 02.03.03-00-033/09-04) in Cracow whose resources were used 
to carry out all the numerical calculations for this project. Thanks also to 
Mariusz Witek and Mi\l osz Zdyba\l{} for their help with CC1 and Hendrik Hoeth
for his help with \Professor. This work was funded in part 
by the Lancaster-Manchester-Sheffield
Consortium for Fundamental Physics under STFC grant ST/J000418/1 and in
part by the 
MCnetITN FP7 Marie Curie Initial Training Network PITN-GA-2012-315877.
\vspace{4ex}
\appendix
\newpage
\section{Tuning details}
\label{Appendix}
In this appendix we briefly show a couple of more technical features
related to our fits.

\subsection{Dependence on the weights}
\label{app:weight}
We argued earlier that an appropriate weight to use for \seff in the
\Professor fits is 132, giving it the same weight as all the
multiplicity or $p_\perp$ sum data. We have also studied the dependence
of the tune on the weight applied.

\begin{figure*}[htb]
  \begin{center}
  \vspace*{-5ex}
\includegraphics[angle=90,width=0.7\textwidth]{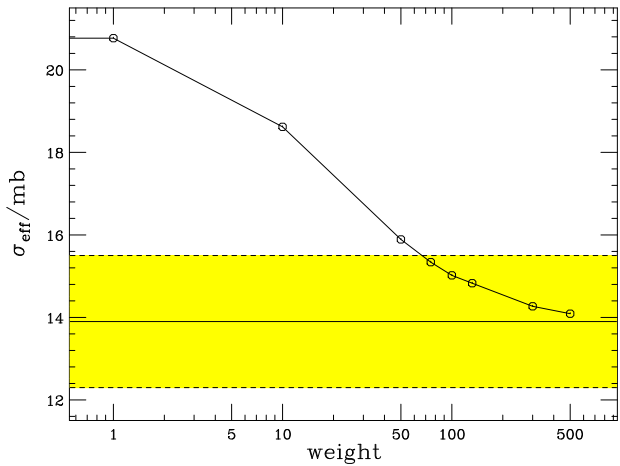}  \end{center}
  \vspace{-10ex}
  \caption{The value of \seffmodel against the weight applied to the
    \seff data in the fit. The horizontal band shows the measured value
    of \seff.}
  \label{fig:weight}
\end{figure*}

In Fig.~\ref{fig:weight} we show the best fit value of \seffmodel as a
function of the weight applied. We see that the weight needs to be at
least about 70 to bring the value of \seffmodel within 1$\sigma$ of its
measured value, and that increasing it significantly beyond about 250
has almost no further effect. We conclude that 132 is a sensible value for
the weight, but that the results are not critically dependent on its
precise value, within a factor of two.

The fact that the UE results are so similar between the ``\textit{No
  \seff in fit}'' fit, which effectively has weight${}=0$, and the \EEvCTEQ
fit, with weight${}=132$, is a confirmation that they do not depend
significantly on the weight applied.

\subsection{Stability of Professor predictions}
\label{app:stability}
\begin{figure*}[htb]
  \includegraphics[width=0.48\textwidth]{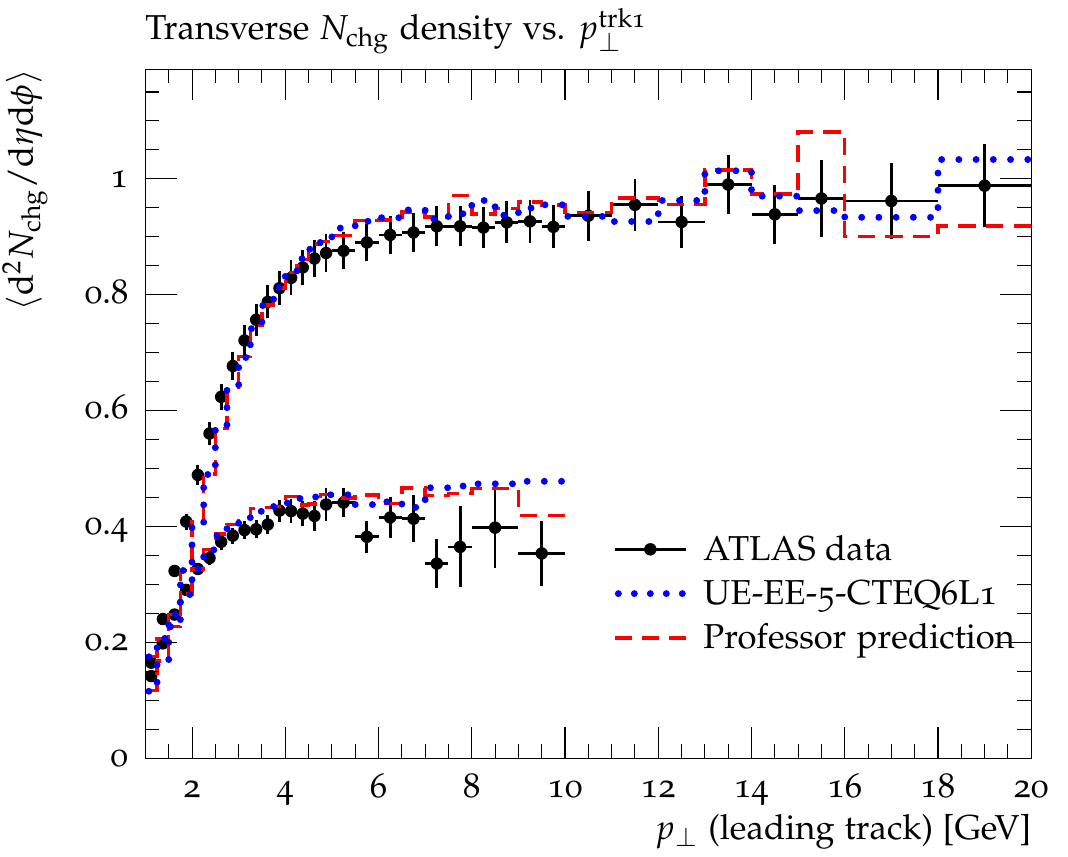}
  \includegraphics[width=0.48\textwidth]{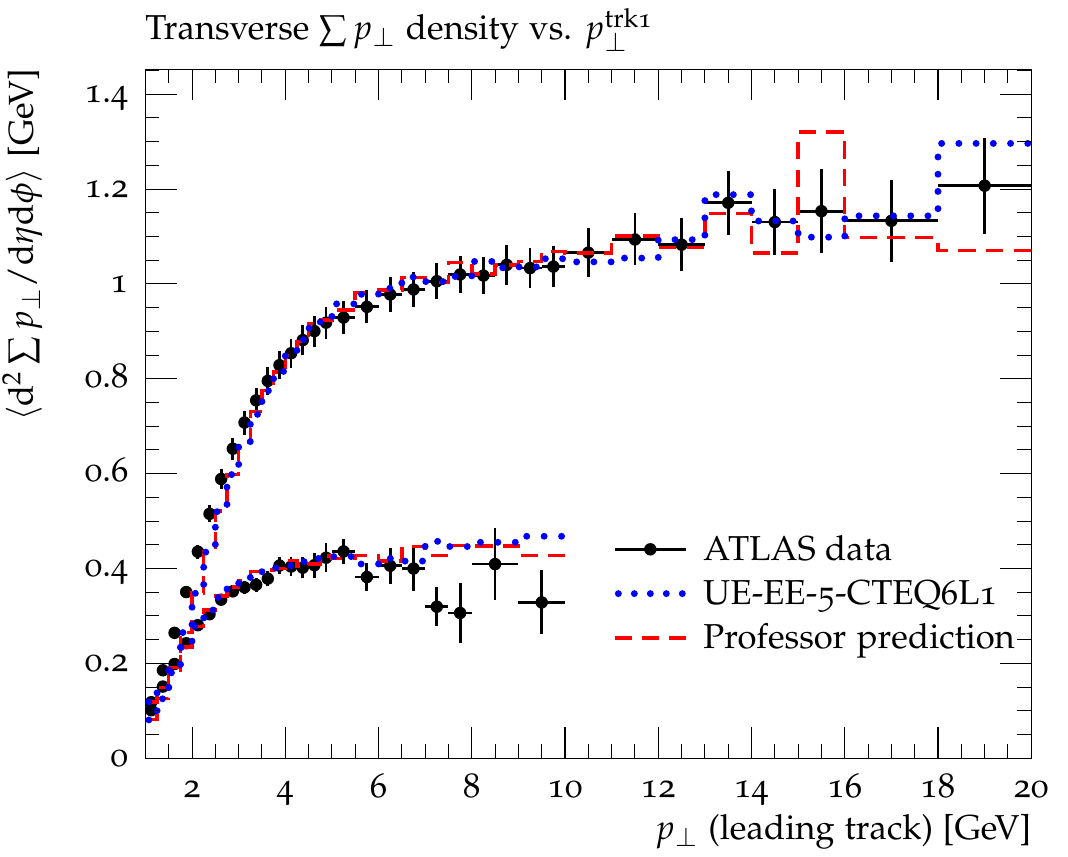}
  \vspace*{-1ex}
    \caption{ATLAS data and model fits at ${900}$~\GeV\ (lower sets) and
      ${7}$~\TeV\ (upper sets), showing the mean multiplicity density
      (left panel) and mean scalar $p_{\perp}$ sum (right panel) of the
      stable charged particles in the ``transverse'' area as a function
      of $\ptlead$, showing the results of the tune \EEvCTEQ and the
      Professor prediction for comparison.}
  \vspace*{-1ex}
  \label{appfig:LHC}
\end{figure*}

\begin{figure*}[htb]
  \includegraphics[width=0.48\textwidth]{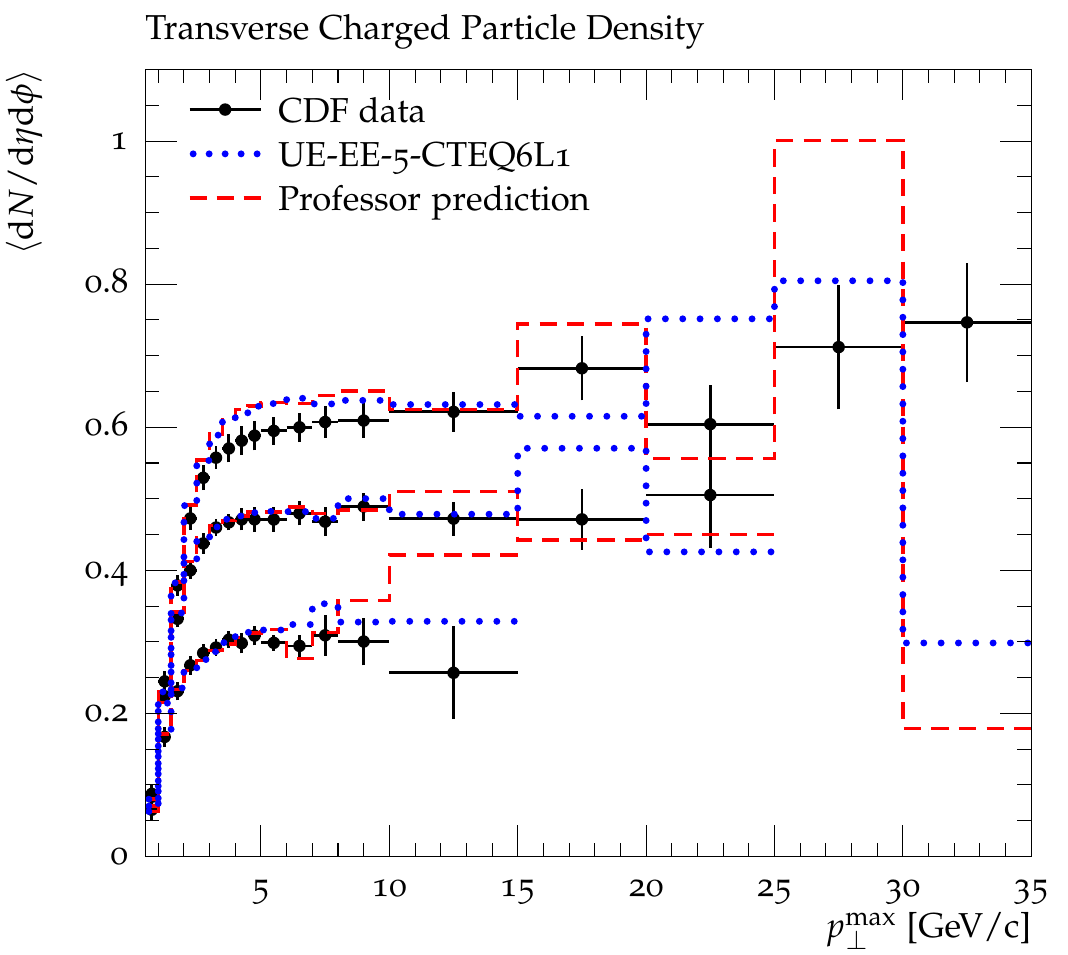}
  \includegraphics[width=0.48\textwidth]{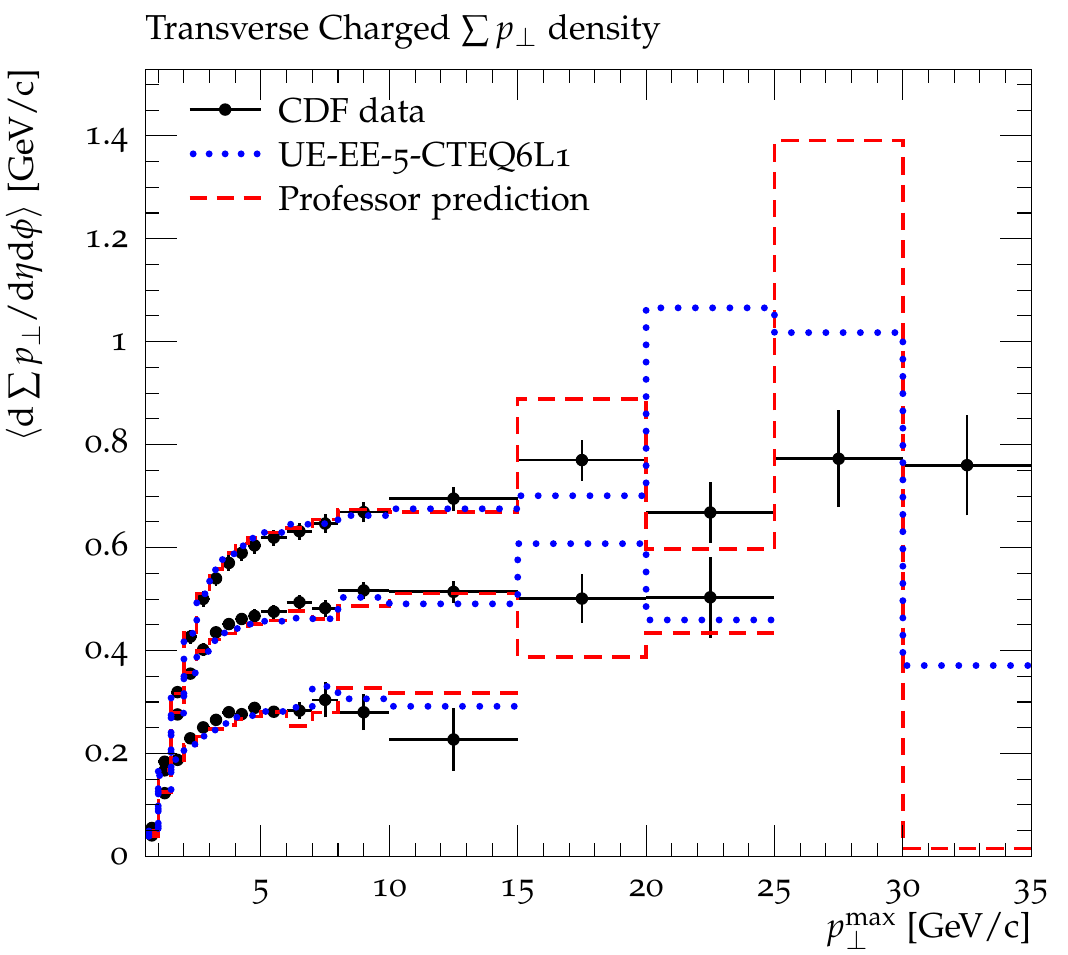}
  \vspace*{-1ex}
    \caption{CDF data and model fits at ${300}$~\GeV\ (lowest sets),
      ${900}$~\GeV\ (middle sets) and ${1960}$~\GeV\ (top sets),
      showing the mean multiplicity density (left panel) and mean scalar
      $p_{\perp}$ sum (right panel) of the stable charged particles in
      the ``transverse'' area as a function of $\ptlead$, showing the
      results of the tune \EEvCTEQ and the Professor prediction for
      comparison.}
  \vspace*{-2ex}
  \label{appfig:TVT}
\end{figure*}

All of our tuning has been performed using \Professor's parametrization
of \herwig{++}'s results as a function of its parameters. As a
cross-check that this parametrization is reliable, we show in
Figs.~\ref{appfig:LHC} and~\ref{appfig:TVT} the comparison of Monte
Carlo runs with the \EEvCTEQ parameter set with \Professor's prediction
of them. They can be seen to agree very well.

We also mention that we only fitted to the regions of the plots where
the experimental and Monte Carlo statistical errors are small emough
that fluctuations are not significant. Specifically, we use the data
with $\ptlead < 5$~\GeV\ at 300~\GeV, $\ptlead < 6$~\GeV\ at 900~\GeV,
$\ptlead < 7$~\GeV\ at 1960~\GeV\ and $\ptlead < 10$~\GeV\ at 7~\TeV. By
eye, one can see that the results are reliable beyond the region to
which they are fit.



\bibliographystyle{JHEP}	
\bibliography{references}



\end{document}